\documentstyle[epsfig,11pt,newpasp,twoside]{article}
\def\edcomment#1{\iffalse\marginpar{\raggedright\sl#1\/}\else\relax\fi}
\reversemarginpar

\begin{document}
\title{Far-Infrared Source Counts and the Diffuse Infrared Background}

\author{J. L. Puget, G. Lagache} 
 
\affil{Institut d'Astrophysique Spatiale, Orsay, 91405, France}

\begin{abstract}
The cosmic far-infrared background is now well measured from 140 $\mu$m
to 1 mm. Uncertainties remain at 100 $\mu$m (and even more at 60 $\mu$m).  These are
dominated by limitations of the zodiacal model. The nature of sources dominating the
background near its maximum are beginning to be studied through deep surveys carried out with
ISOPHOT.
These surveys show very steep number counts and fluctuations
of the background due to unresolved sources.
\end{abstract}

\section{The Cosmic Far-InfraRed Background (CFIRB)}

\subsection{Determination of the CFIRB above 100 $\mu$m: another Approach}
In the parts of the sky with no molecular clouds
or dense HII regions, the far-IR emission 
can be written as the sum of dust emission associated
with the neutral gas and with the diffuse ionised gas, 
interplanetary dust emission, the CFIRB, and the cosmic microwave background  
and its dipole. In previous studies (Puget et al. 1996; Fixsen et al. 1998; 
Hauser et al. 1998), 
dust emission associated with the ionised gas
which could not be traced independently has been either not subtracted
properly or neglected. 

Lagache et al. (1999) have for the first time detected 
dust emission in the ionised gas 
and shown that the emissivity (which
is the IR emission normalised to unit hydrogen column density)
of dust in the ionised gas is nearly the same as in the
neutral gas. This has consequences for the determination of the CFIRB.
Following this first detection, Lagache et al. (2000) have combined HI and 
WHAM H$_{\alpha}$ data (Reynolds et al. 1998)
with far-IR COBE data in order to derive dust
properties in the diffuse ionised gas as well as to make a proper
determination of the CFIRB.
After careful pixel selection
they describe the 
far-infrared dust emission as a function of the HI
and H$^+$ column densities by:
\begin{equation}
\label{main_EQ}
IR= A \times N(HI)_{20{\rm cm}^{-2}} + B \times N(H^{+})_{20{\rm cm}^{-2}} + C
\end{equation}
where N(HI)$_{20{\rm cm}^{-2}}$ and N(H$^+$)$_{20{\rm cm}^{-2}}$ 
are the column densities normalised 
to 10$^{20}$ hydrogen atoms and ions per square centimeter respectively. The coefficients A, B
and the constant term C are determined simultaneously using regression fits. 
They show that about 25$\%$ of the IR
emission comes from dust associated with the diffuse ionised gas
which is in very good agreement with the first determination
of Lagache et al.~(1999). 
The CFIRB spectrum obtained using this galactic
far-infrared emission decomposition is shown in Figure 1 
together with the CFIRB FIRAS determination
of Lagache et al.~(1999) in the Lockman Hole region. We see very
good agreement between the two spectra, as well with
the Fixsen et al.~(1998) determination (Figure 2).
Fixsen et al.~(1998) used an infrared tracer for the interstellar component 
which is equivalent as long as the dust emission spectrum from the diffuse 
neutral and ionised interstellar gas are very similar. This hypothesis is 
confirmed by the Lagache et al.~(2000) analysis.

\begin{figure}
\centering
\includegraphics[width=.8\textwidth]{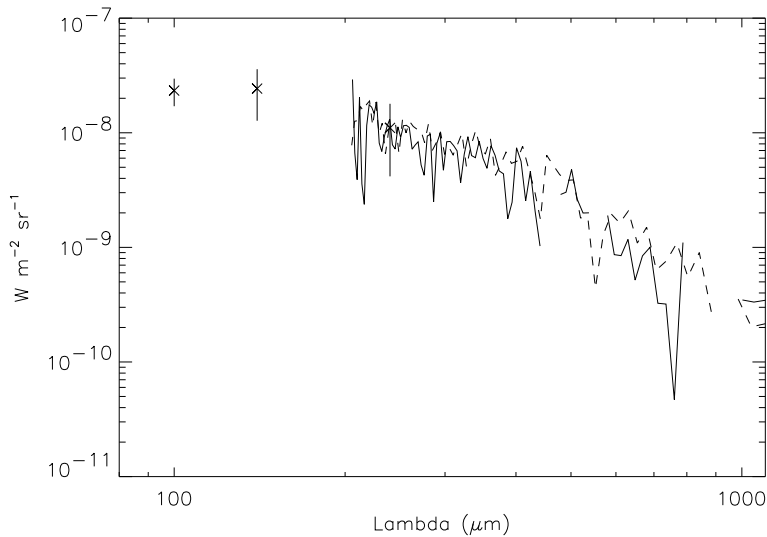}
\caption[]{\label{CFIRB_spec}CFIRB spectra obtained from the decomposition
of the far-infrared sky (continuous line) and determined for the Lockman
Hole region (dashed line) by Lagache et al.~(1999). Also reported
are DIRBE values at 100, 140 and 240 $\mu$m from Lagache et al.~(2000).}
\end{figure}

\begin{figure}
\centering
\includegraphics[width=.8\textwidth]{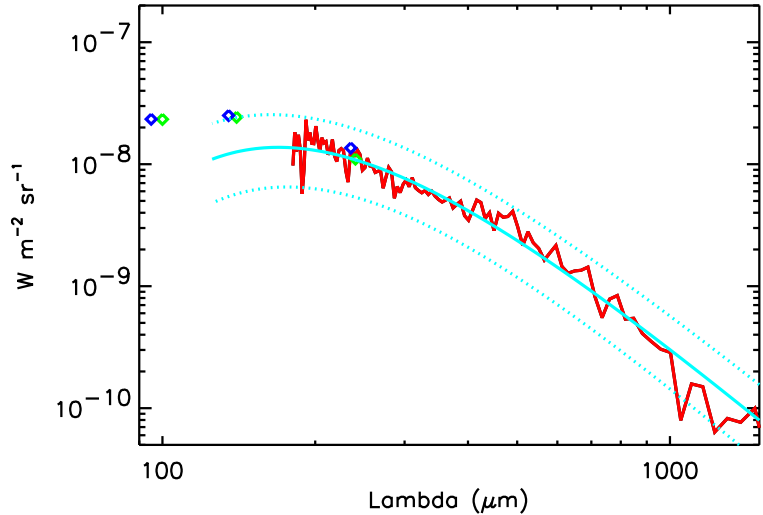}
\caption[]{\label{CFIRB_compa}Comparison of the CFIRB determinations: Lagache
et al. 2000 (jagged continuous line and light diamonds), Hauser et al. 1998 (dark diamonds,
arbitrarily shifted), Fixsen et al.~1998 (dotted and smooth continuous lines)}
\end{figure}

At 140 and 240 $\mu$m, the values obtained for the CFIRB
are 1.13$\pm$0.54~MJy/sr and 0.88$\pm$0.55~MJy/sr respectively. 
For each selected pixel, Lagache et al.~(2000) compute
the residual emission, R = IR - A$\times$N(HI) - B$\times$N(H$^+$).
Uncertainties in the CFIRB have been derived from the width of the histogram
of R (statistical uncertainties derived from the regression analysis are negligible).
The CFIRB values obtained, although much more noisy
(due to the small fraction of the sky used), are in very
good agreement with the determination of Hauser et al.~(1998).
At 140~$\mu$m, the CFIRB value of Lagache et al.~(1999) is smaller 
than that derived here since the assumed WIM (Warm Ionised
gas) dust spectrum was overestimated (the WIM dust spectrum was
very noisy below 200~$\mu$m and the estimated dust
temperature was higher).
At 100~$\mu$m, 
assuming an accurate subtraction of the zodiacal emission,
the far-IR emission decomposition gives:
I$_{CFIRB}$(100)= 0.78$\pm$0.21~MJy/sr. This is the first 
time that two independent gas tracers for the HI and the
H$^+$ have been used to determine the background at 100 
$\mu$m. 
The CFIRB value of 0.78~MJy/sr can be compared to the non-isotropic residual
emission found by Hauser et al.~(1998). The average over three regions
of the residual emission, equal
to 0.73$\pm$0.20~MJy/sr, is in very good agreement with the 
determination of Lagache et al.~(2000).\\

So we see, using different approaches, that we are now converging
on the shape and level of the CFIRB above 100~$\mu$m.

\subsection{Where are the Main Uncertainties?}
There are two main uncertainties in the CIRB (Cosmic InfraRed
Background) 
determination in the IR and sub-mm domain: 
(1) Dust emission associated with the ionised gas and 
(2) Zodiacal emission.\\

If the spectrum of dust associated 
with the ionised and neutral gas is the same, a CFIRB 
determination using color ratios should give the same values
as other methods implying independent HI and H$_{\alpha}$
gas tracers.
However, one has to note that methods based on
the intercept of the far-IR/HI correlation
for the determination of the CFIRB 
are dangerous. For example, for the parts of the sky selected 
by Lagache et al.~(2000), this intercept is about 0.91~MJy/sr, which 
is quite different from the value of the CFIRB. \\

The zodiacal emission was obtained by Kelsall et al.~(1998)
relying on its time variability. Other models have been built 
for the zodiacal emission which is critical to a determination of the CIRB in 
the near infrared (Wright 2000; Finkbeiner et al.~2000). The accuracy of these models 
can be estimated by comparing the residuals observed at wavelengths were the 
zodiacal emission is maximum (12 and 25 $\mu$m). For example, on the one hand, the Kelsall et al.~(1998)
model leads to residuals of about $4.7\times 10^{-7}$ Wm$^{-2}$sr$^{-1}$
at these wavelengths. 
On the other hand, upper limits on the CIRB have been established in 
this wavelength range using TeV gamma rays from extragalactic sources 
at the level of 10$^{-8}$ Wm$^{-2}$sr$^{-1}$ (e.g., Biller et al.~1995). This is not in
contradiction 
with the estimate of uncertainties in the Kelsall et al.~zodiacal model by Hauser et al.~(1998), 
who give only upper limits for the CIRB. Nevertheless this result,
together with the fact that the residuals follow a zodiacal emission
spectrum, suggests that this best model underestimates 
the zodiacal emission as the other contributions (instrumental and 
interstellar) are significantly smaller. We can thus make a conservative estimate 
of the amount of zodiacal
emission which is not removed by this model at 12 and 25 $\mu$m to be  
$4\times 10^{-7}$ Wm$^{-2}$sr$^{-1}$. The amounts not removed at 60 and 100 $\mu$m are
thus $4\times 10^{-8}$ Wm$^{-2}$sr$^{-1}$ and $8.4\times 10^{-9}$ Wm$^{-2}$sr$^{-1}$
(using the Kelsall et al.~smooth high latitude zodiacal cloud
color ratios). 
This reduces the CIRB at 100~$\mu$m from 0.78 to 0.50 MJy/sr. At 60~$\mu$m
the extra zodiacal emission to be removed ($4\times 10^{-8} $ Wm$^{-2}$sr$^{-1}$) 
is comparable to the residuals, and thus no meaningful value can be obtained on the 
CIRB at this wavelength.

\section{Resolving the CFIRB at 170~$\mu$m}

\subsection{The FIRBACK Deep Survey}
FIRBACK is a survey of 4 square degrees in 3 high galactic latitude fields, 
chosen to have as low a HI column-density as possible, 
typically $N_{HI} \simeq 10^{20} {\rm cm}^{-2}$, and as much multiwavelength coverage as
possible.
Observations were carried out with the ESA Infrared Space Observatory, ISO, (Kessler et al.~1996) 
using the ISOPHOT photometer (Lemke et al.~1996) with the 
C200 camera and C$_{160}$ broadband filter centered at $\lambda = 170\, \mu$m. 
A detailed description of the reduction, data processing, and calibration will be discussed in 
Lagache \& Dole (2001); the analysis of the complete survey is
discussed in Dole et al.~(2001).\\

Eighty-six sources are detected above the sensitivity limit of the survey (200~mJy, 5$\sigma $). 
The number of sources detected above 120 mJy ($ 3 \sigma $) is 235. The first result of this
survey is the high number of sources observed when compared to 
no, or moderate, evolution models for infrared galaxies. Extensive simulations were done to
establish noise properties, incompleteness (80$ \% $ at 200~mJy) and Eddington bias; the cumulative
number counts were then established from this catalog.  
These number counts are extremely steep, $ N(>S) \propto S^{-2.2} $, indicating very strong 
cosmological evolution of infrared galaxies.
This is illustrated in Figure 3(taken from Dole et al. 2000). Two models from Guiderdoni et al.~(1998)
are shown. It can be noticed that the semi-empirical model ``E'', which was built to explain the 
infrared background, fits the counts well but falls slightly short of explaining the steepness of
the counts.\\

Number counts in a small field (0.25 square degrees) have already been published (Puget et al.
1999). This field was observed as a ``feasibility demonstration" of the FIRBACK survey at a time when
the limitations to weak source detection with ISOPHOT were not yet well understood. The counts given in
Puget et al.~(1999)
are compatible with the counts shown in Fig. 3.\\

Integrating these counts for all sources detected above 120 mJy (3$\sigma$) gives a brightness of 
1.6 nW m$^{-2}$ sr$^{-1}$ or 8$\%$ of the CIRB at 170~$\mu$m. If we extrapolate the counts 
to weaker fluxes with a slope -2, we find that the background is fully accounted for by 
sources brighter than about 10~mJy. As we expect the counts to flatten in a progressive way
we predict that the background at 170~$\mu$m is likely to be dominated by sources of a few~mJy.

\begin{figure}
\centering
\includegraphics[width=.8\textwidth]{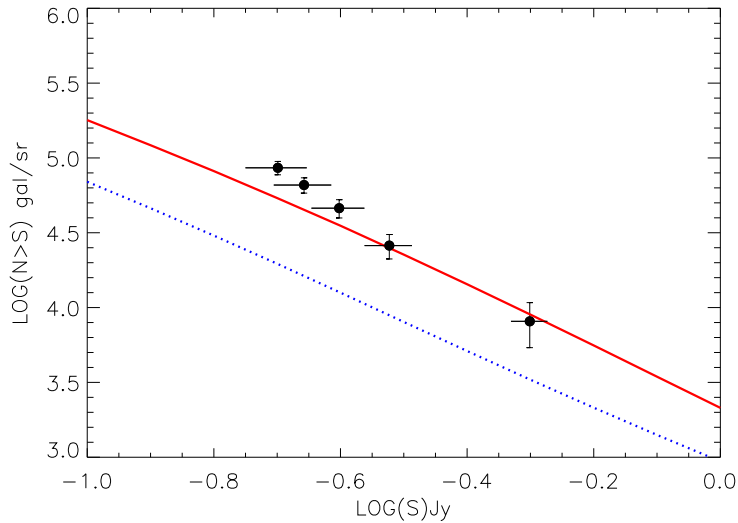}
\caption[]{\label{FIRBACK_count} FIRBACK galaxy counts superposed on
the Guiderdoni et al.~(1998) models ``A'' (dotted line) and ``E'' (continuous line)(from Dole et al. 2000).}
\end{figure}

Considering the rather low angular resolution allowed by a 60~cm telescope at this wavelength, 
the identification of these sources with optically detected sources is difficult. 
Starburst galaxies are expected to be radio sources with a well defined ratio of infrared to radio
flux.
Radio surveys have been carried out in the FIRBACK fields with the VLA for the two 
northern fields (Ciliegi et al.~1998) and the Australian Telescope for the southern field.  
Furthermore, 15~$\mu$m observations have been carried out with the ISOCAM instrument
aboard ISO (Oliver et al.~2000).
These two sets of observations are the best tools for initiating the identification process when 
counterparts are available --- which is the case for about 50$\%$ of our sources. Confirming the
identification is best done with ground based millimeter or submillimeter observations with SCUBA, 
CSO or the IRAM 30 meter telescope. The last step is optical identification using the radio
positions if a millimeter or submillimeter detection has confirmed the FIRBACK-radio tentative
identification. 
This process is frustratingly slow.
On the one hand, a substantial fraction of the sources are relatively nearby 
moderate starburst galaxies with bright optical counterparts 
(redshift less than 0.3, far infrared luminosities of a few 10$^{11}$ L$_{\odot}$).
On the other hand, a small fraction of the sources have very weak or no optical counterpart and
are likely to
be associated with distant (redshift larger than 1) ultraluminous starburst or dust enshrouded
AGNs.
 
\subsection{The Lockman Hole Survey}
A survey of comparable depth has been carried out in the Lockman hole (Kawara et al.~1998; 
Matsuhara et al.~2000). The number counts at 170~$\mu$m are in excellent agreement with the 
counts from the FIRBACK survey. It should be noticed that the Lockman hole is one of the fields
with the lowest cirrus content. The agreement in the number counts confirms that, for point sources,
the contamination of the extragalactic source counts by small scale cirrus structure is negligible.
This is also in agreement with the conclusions of the analysis reported in the next section,  
which discusses the fluctuations of the background. The difference in the power spectra of
the cirrus component and of the extragalactic component leads to a negligible cirrus contribution 
to the power spectrum at the highest spatial frequencies explored by this survey.   

\section{The 170~$\mu$m CFIRB Fluctuations}

\subsection{Why Search for the CFIRB Fluctuations?}

The CFIRB is taken to consist of sources with number counts as a function of flux which can
be represented, for the present discussion, by a simple power law:
\begin{equation}
N(>S)= N_0 \left( \frac{S}{S_0} \right)^{- \alpha}
\end{equation}
Obviously, these number counts need to flatten at low fluxes
to insure a finite value of the background. Thus, we assume that $\alpha$=0 
for $S<S^{\ast}$, where $S^{\ast}$ is the flux where the counts flatten.\\

For the simple euclidian case ($\alpha$=1.5), the CFIRB integral 
is dominated by sources near $S^{\ast}$ and
its fluctuations are dominated by sources which are just below the detection
limit $S_0$. It is well known that strong cosmological
evolution, associated with a strong negative K-correction, could lead to a very steep
number count distribution (see, for example, Guiderdoni et al.~1998 and Franceschini et al.~1998). 
The present far-IR observations indeed show a very steep slope of
$\alpha$=2.2 (Dole et al.~2000). In this case,
the CFIRB integral is still dominated by sources near $S^{\ast}$, but
its fluctuations are now also dominated by sources close to 
$S^{\ast}$. Thus, it is essential to study the 
extragalactic background fluctuations, which are likely to have a substantial
contribution from sources with a flux comparable to those dominating the
CFIRB intensity (sources that are not just below the sensitivity
limit).\\

To see if we can detect the CFIRB fluctuations, we need wide field
far-IR observations with high angular resolution and very high signal to noise ratio.
The FIRBACK project fulfills all these conditions. 
To search for CFIRB fluctuations, we have first used the so-called
``Marano 1'' field, for which we obtained 4 independent coadded
maps which allow us to properly determine the instrumental noise
(Lagache \& Puget~2000).
In this field, we have a signal to noise ratio of about 300
and we detect 24 sources 
that we remove from the original map.
We first carry out an analysis on this field, and then extend it to the other FIRBACK fields.\\

\subsection{CFIRB Fluctuation Mean Level}
FIRBACK maps show background fluctuations which consist of two components that we want to
separate, 
galactic cirrus fluctuations and, if present, extragalactic fluctuations. \\

Our separation of the extragalactic and galactic fluctuations
is based on a power spectrum decomposition. This method
allows us to discriminate between the two components using the statistical
properties of their spatial behaviour. Fig.~4 
shows the power spectrum of the ``Marano~1'' field.
In the plane of the detector, the power spectrum measured on the map 
can be expressed in the form:
\begin{equation}
P_{map}= P_{noise} + (P_{cirrus} + P_{sources}) \times W_k 
\label{Eq_Pk}
\end{equation}
where  $P_{noise}$ is the instrumental noise power spectrum measured
using the 4 independent maps of the Marano~1 field (Lagache \& Puget~2000), 
$P_{cirrus}$ and $P_{sources}$ are the cirrus and
unresolved extragalactic source power spectra respectively, 
and $W_k$ is the power spectrum of the point spread function, PSF. For our analysis,
we remove $P_{noise}$ from $P_{map}$.\\

\begin{figure}
\centering
\includegraphics[width=.8\textwidth]{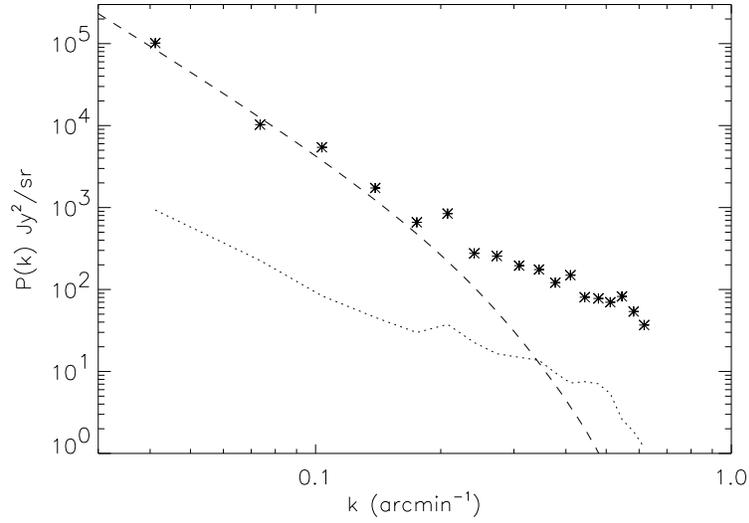}
\caption[]{\label{fluc_Marano}Power spectrum of the source subtracted ``Marano 1'' field
($\ast$).
The instrumental noise power spectrum (dotted line) has been subtracted. The dashed line
represents the cirrus power spectrum multiplied by the power spectrum of the PSF.}
\end{figure}

We know from previous work that the cirrus far-infrared 
emission power spectrum has a steep slope, $P_{cirrus} \propto k^{-3}$  
(Gautier et al.~1992; Kogut et al.~1996; Herbstmeier et al.~1998; Wright 1998; 
Schlegel et al.~1998; Miville-Desch\^enes et al.~in preparation). 
These observations cover the relevant spatial frequency range and have been recently extended
up to 1 arcmin using HI data of very diffuse regions (Miville-Desch\^enes~1999). 
The extragalactic
component is unknown but certainly much flatter (see the discussion in Lagache \&
Puget~2000). 
We thus conclude
that the steep spectrum observed in our data at $k<$0.15 arcmin$^{-1}$ 
(Fig. 4) can only be due to cirrus emission. 
The break in the power spectrum at $k\sim$ 0.2 arcmin$^{-1}$ is 
very unlikely to be due to the cirrus emission itself which is 
known not to exhibit any preferred scale (Falgarone et al.~1998). 
Thus, the normalisation of our cirrus power spectrum $P_{cirrus}$ is directly determined
using the low frequency data points and assuming a $k^{-3}$ dependence.\\

\begin{figure}
\centering
\includegraphics[width=.8\textwidth]{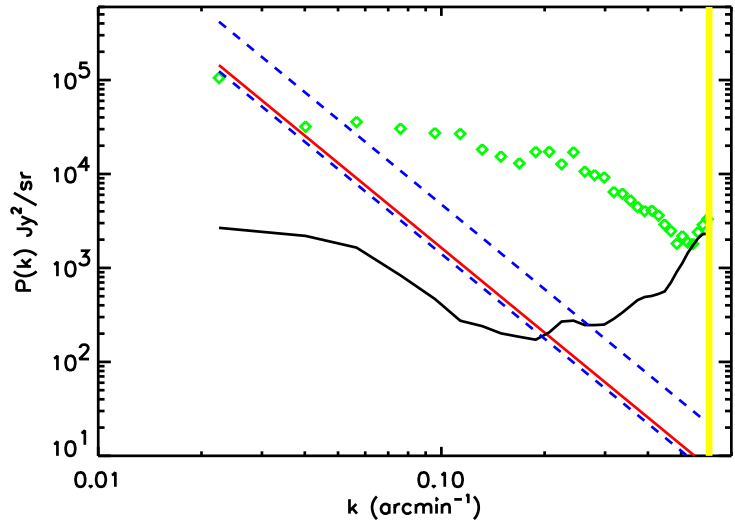}
\caption[]{\label{fluc_N2_all}Power spectrum of the FIRBACK/ELAIS N2 field
($\diamond$). The straight continuous and dashed lines, respectively, are the best fit cirrus power spectrum and the spectrum deduced from Miville-Desch\^enes et al (in preparation).  The continuous curve shows the detector noise.}
\end{figure}

We clearly see in Figure~4 an excess over $P_{cirrus}$ between
$k$=0.25 and 0.6~arcmin$^{-1}$, which is more than a factor of 10
at $k$=0.4~arcmin$^{-1}$. Any reasonable power law spectrum
for the cirrus component multiplied by the footprint
leads, as can be easily seen in Fig.~4, to a very steep
spectrum at spatial frequency $k>$0.2 arcmin$^{-1}$, 
which is very different from the observed spectrum.
Moreover, the excess is more than
10 times larger than the measured instrumental noise power spectrum.
Therefore, as no other major source of fluctuations
is expected at this wavelength, the large excess observed between
k=0.25 and 0.6~arcmin$^{-1}$ is interpreted as due
to unresolved extragalactic sources.\\

\begin{figure}
\centering
\includegraphics[width=.8\textwidth]{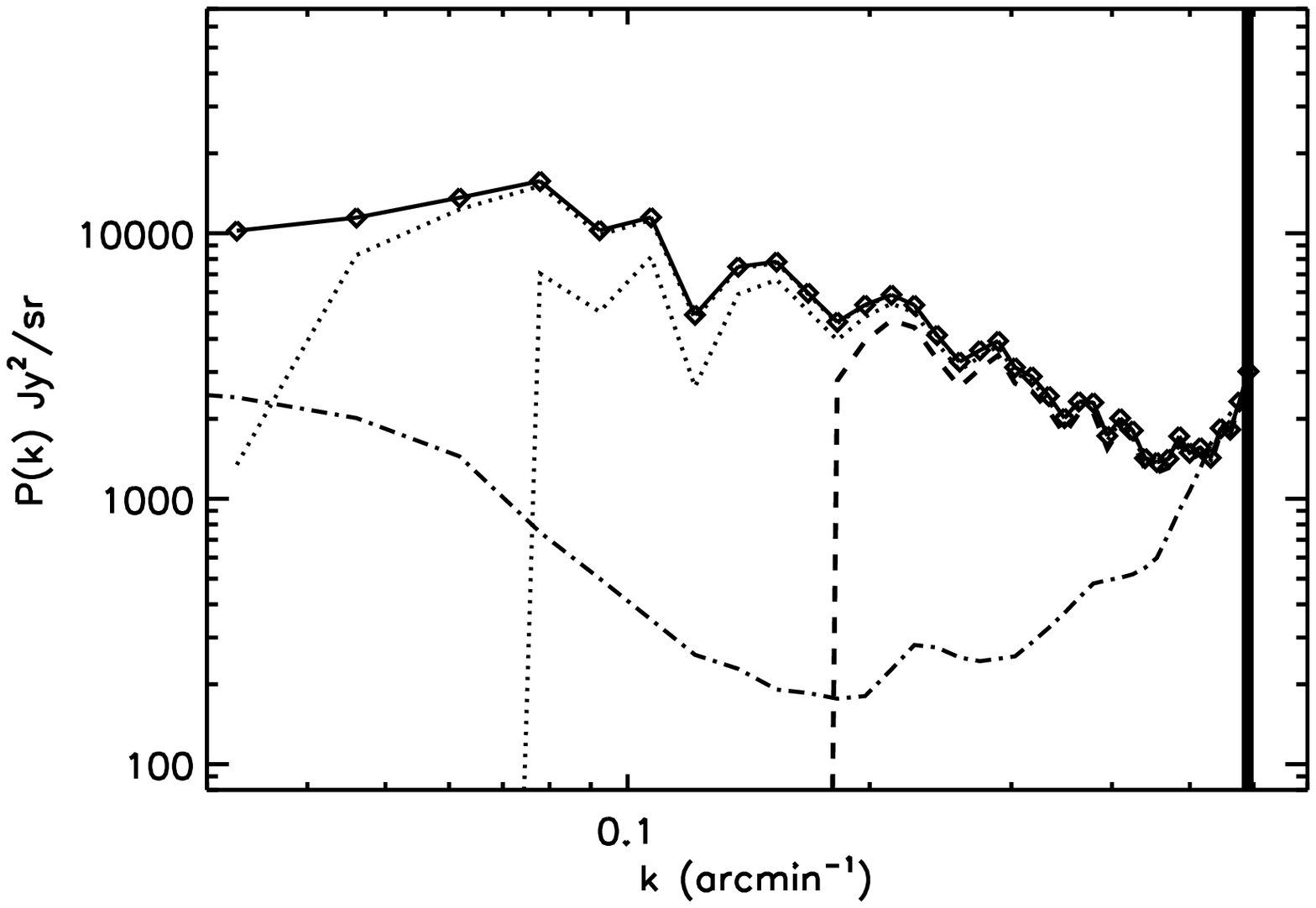}
\caption[]{\label{fluc_N1}Power spectrum of the FIRBACK N1 extragalactic component.  The continuous curve with $\diamond$ symbols represents the residual after the best-fit cirrus power spectrum has been subtracted from the power spectrum of the observational data in this field.  The dotted lines represent the residuals after subtraction of the two power spectra of Miville-Desch\^enes et al., analogous to the two shown in Figure 5 for the N2 field.  The dot-dashed line is the detector noise spectrum.  The dashed line is the residual extragalactic power spectrum obtained by subtracting 10 times the best-fit cirrus power spectrum from the directly observed power spectrum for this field.}
\end{figure}
\begin{figure}
\centering
\includegraphics[width=.8\textwidth]{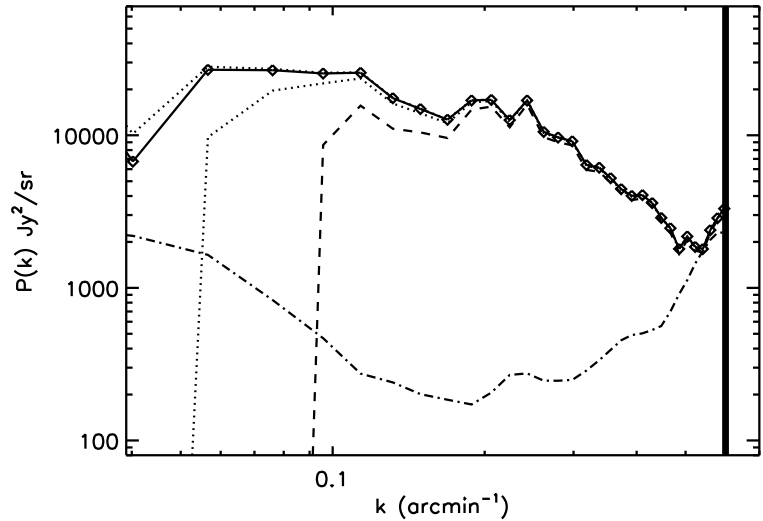}
\caption[]{\label{fluc_N2} Same as Fig. 6 but for the FIRBACK/ELAIS N2 field.}
\end{figure}

The Marano 1 field cannot be used to
constrain the clustering of galaxies due to its rather small size.
However, the extragalactic source power spectrum mean level can be determined.
We obtain $P_{sources}$~=~7400~Jy$^2$/sr, which is in very good
agreement with the spectrum predicted by Guiderdoni et al.~(1997).
This gives CFIRB rms fluctuations around 0.07~MJy/sr (for
a range of spatial frequencies up to 5~arcmin$^{-1}$). 
These fluctuations are at the $\sim$9 percent level,
which is very close to the predictions of Haiman \& Knox (2000).\\

The same analysis can be applied to the other and larger FIRBACK
fields (N1 and N2). Fig.~5 shows the fluctuation
power spectrum ($P_{map}$) obtained for the FIRBACK/ELAIS N2 field in the plane of the sky. 
Again, we clearly see a large excess over $P_{cirrus}$.\\

From Eq.~3, we deduce: 
$$P_{sources}= (P_{map} - P_{noise}) / W_k - P_{cirrus}.$$
We see in the N1 and N2 fields (Fig.~6 and 7) an 
increase by a factor of 5 to 10 of $P_{sources}$ from small to large 
scales. This increase cannot be due to cirrus: for example, the removal of 
10 times the cirrus $P_k$ leads to the same increase (dashed line in the two figures). So, 
it seems that we have with FIRBACK the first indication of IR sources clustering.
Although the present analysis cannot be conclusive on this question, it clearly
shows that the FIRBACK survey has the required sensitivity and is free enough from
contamination to deal with this question.  \\

If we remove from the N1 and N2 maps all detected sources ($S>$200~mJy, 5$\sigma$), 
we obtain CIB fluctuations which are compatible with a Poissonian 
distribution (Fig.~8). We need in fact in this case an independent cirrus
template (like that for the HI gas) to remove the galactic contribution more correctly
and see if correlations are present. \\
\begin{figure}
\centering
\includegraphics[width=.8\textwidth]{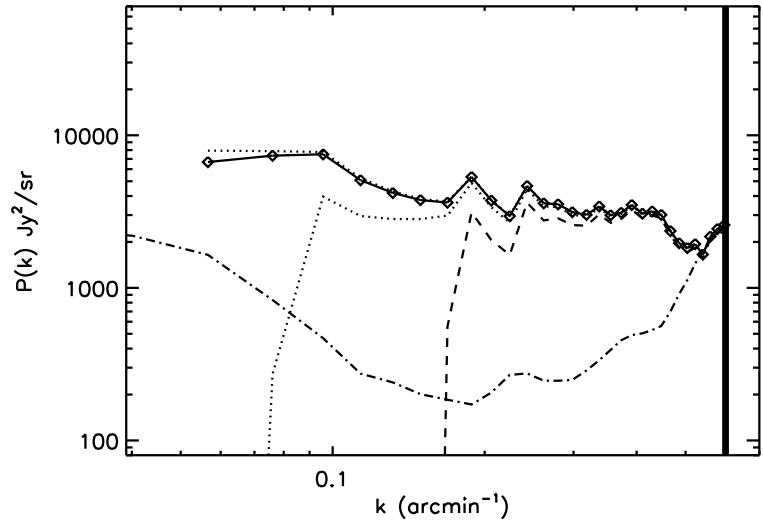}
\caption[]{\label{fluc_N2_200}Power spectrum of the FIRBACK/ELAIS N2 unresolved
extragalactic component (sources with flux $S>$200~mJy have been
removed from the maps.) Line symbols are
the same as in Fig. 6.}
\end{figure}

In summary, we have seen that the fluctuations probe the CFIRB dominant source
population. We detect the clustering of the detected FIRBACK sources. 
On the contrary, the dominant contribution
to the CIB fluctuations seems to be compatible with a Poissonian distribution.\\

It is important to note that the level
of the extragalactic fluctuations derived from FIRBACK fields
is in very good agreement with the Matsuhara et al.~(2000, and this conference) study in the
Lockman Hole region, where the interstellar dust contamination is negligible.
The presence of correlated extragalactic fluctuations should be studied
using all the extragalactic ISOPHOT fields and independent cirrus tracers.

\vspace{5mm}
\centerline{Discussion}
\vspace{5mm}

Martin Harwit:  While the SCUBA sources might be predominantly at high redshifts, and cannot
be ruled out as being at high $z$, the background is so low there that these galaxies are not major
factors as far as production of heavy elements or massive star formation is concerned.  Otherwise
the abundance of chemical elements at high redshifts would be seen to be much higher.

Jean-Loup Puget: Yes. In the model I presented the bulk of the energy comes from redshifts
around $z = 1$, as do the heavy elements.

Ray Norris:  At higher redshift, metallicity will decrease, so presumably the dust fraction will
decrease.  To what extent can this account for your observed luminosity evolution?

Puget:  The luminosity function I showed is a purely empirical one fitting both the background
and the counts at 15, 170, and 850\,$\mu$m.  The dust contribution, which should decrease at
large $z$, has to be combined with morphology of the galaxies.  Local starbursts show much
larger extinction than the average one.

\end{document}